\documentclass[sigconf]{acmart}

\usepackage{amsmath}
\usepackage{graphicx}
\usepackage{here}
\PassOptionsToPackage{hyphens}{url}
\usepackage{hyperref} 
\usepackage{url} 
\usepackage{textcomp} 
\usepackage{multicol}
\usepackage{subcaption}
\usepackage{comment}
\usepackage{booktabs,dcolumn}
\usepackage{multirow}
\usepackage{siunitx} 
\usepackage{fancyhdr}			
\usepackage{arydshln}
\usepackage{xspace}

\usepackage{multirow}

\usepackage{soul, xcolor}
\usepackage{etoolbox} 
\setstcolor{red}

\usepackage{array}
\usepackage{booktabs}
\usepackage{longtable}
\usepackage{tabularx}

\AtBeginEnvironment{quote}{\itshape}
\AtBeginDocument{%
\providecommand\BibTeX{{%
\normalfont B\kern-0.5em{\scshape i\kern-0.25em b}\kern-0.8em\TeX}}}

\renewenvironment{quote}[1][0.04\linewidth]
{\list{}{\leftmargin=#1\rightmargin=#1}\item\relax}{\endlist}

\usepackage{color}
\definecolor{orange}{RGB}{255,127,0}
\definecolor{limegreen}{RGB}{50, 205, 50}
\definecolor{violet}{RGB}{148,0,211}

\newcolumntype{L}[1]{>{\raggedright\let\newline\\\arraybackslash\hspace{0pt}}m{#1}}
\newcolumntype{C}[1]{>{\centering\let\newline\\\arraybackslash\hspace{0pt}}m{#1}}
\newcolumntype{R}[1]{>{\raggedleft\let\newline\\\arraybackslash\hspace{0pt}}m{#1}}

\newif\ifCOMMENTS
\COMMENTSfalse

\ifCOMMENTS

\newcommand{\revise}[2]{\textcolor{red}{\sout{#1}}~\textcolor{blue}{#2}}
\newcommand{\revision}[1]{{\color{blue} #1}}

\else

\newcommand{\revise}[2]{#2}
\newcommand{\revision}[1]{#1}

\fi

\newcommand{\camera}[2]{#2}

\makeatletter
\def\Hline{
\noalign{\ifnum0=`}\fi\hrule \@height 4.\arrayrulewidth \futurelet
\reserved@a\@xhline}
\makeatother

\def\oursys{\textit{Robo-Blocks}\xspace}

\AtBeginDocument{%
  \providecommand\BibTeX{{%
    Bib\TeX}}}

\copyrightyear{2025}
\acmYear{2025}
\setcopyright{cc}
\setcctype{by}
\acmConference[CHI '25]{CHI Conference on Human Factors in Computing Systems}{April 26-May 1, 2025}{Yokohama, Japan}
\acmBooktitle{CHI Conference on Human Factors in Computing Systems (CHI '25), April 26-May 1, 2025, Yokohama, Japan}\acmDOI{10.1145/3706598.3713718}
\acmISBN{979-8-4007-1394-1/25/04}

\acmSubmissionID{1323}

\begin{document}

\title{Bridging Generations using AI-Supported Co-Creative Activities}
\author{Callie Y. Kim}
\email{cykim6@cs.wisc.edu}
\orcid{0009-0001-4195-8317}
\affiliation{%
  \institution{Department of Computer Sciences\\University of Wisconsin--Madison}
  \city{Madison}
  \state{Wisconsin}
  \country{USA}
}

\author{Arissa J. Sato}
\email{asato@wisc.edu}
\orcid{0000-0002-1103-8050}
\affiliation{%
  \institution{Department of Computer Sciences\\University of Wisconsin--Madison}
  \city{Madison}
  \state{Wisconsin}
  \country{USA}
}

\author{Nathan Thomas White}
\email{ntwhite@wisc.edu}
\orcid{0009-0000-9414-9647}
\affiliation{%
  \institution{Department of Computer Sciences\\University of Wisconsin--Madison}
  \city{Madison}
  \state{Wisconsin}
  \country{USA}
}

\author{Hui-Ru Ho}
\email{hho24@cs.wisc.edu}
\orcid{0009-0000-3701-2521}
\affiliation{%
  \institution{Department of Computer Sciences\\University of Wisconsin--Madison}
  \city{Madison}
  \state{Wisconsin}
  \country{USA}
}

\author{Christine P Lee}
\email{cplee5@cs.wisc.edu}
\orcid{0000-0003-0991-8072}
\affiliation{%
  \institution{Department of Computer Sciences\\University of Wisconsin--Madison}
  \city{Madison}
  \state{Wisconsin}
  \country{USA}
}

\author{Yuna Hwang}
\email{yunahwang@cs.wisc.edu}
\orcid{0000-0001-7726-8003}
\affiliation{%
  \institution{Department of Computer Sciences\\University of Wisconsin--Madison}
  \city{Madison}
  \state{Wisconsin}
  \country{USA}
}

\author{Bilge Mutlu}
\email{bilge@cs.wisc.edu}
\orcid{0000-0002-9456-1495}
\affiliation{%
  \institution{Department of Computer Sciences\\University of Wisconsin--Madison}
  \streetaddress{Department of Computer Sciences, University of Wisconsin--Madison}
  \city{Madison}
  \state{WI}
  \country{United States}
}

\renewcommand{\shortauthors}{Kim et al.}

\begin{abstract}
Intergenerational \camera{collaboration}{co-creation using technology} between grandparents and grandchildren can be challenging due to differences in technological familiarity. AI has emerged as a promising tool to support \camera{collaborative}{co-creative} activities, offering flexibility and creative assistance, but its role in facilitating intergenerational \camera{collaboration}{connection} remains underexplored. In this study, we conducted a user study with 29 grandparent-grandchild groups engaged in AI-supported story creation to examine how AI-assisted co-creation can \camera{facilitate}{foster} \camera{collaboration}{meaningful intergenerational bonds}. Our findings show that grandchildren managed the technical aspects, while grandparents contributed creative ideas and guided the storytelling. AI played a key role in structuring the activity, facilitating brainstorming, enhancing storytelling, and balancing the contributions of both generations. The process fostered mutual appreciation, with each generation recognizing the strengths of the other, leading to an engaging and cohesive co-creation process. We offer design implications for integrating AI into intergenerational co-creative activities, emphasizing how AI can enhance \camera{collaboration}{connection} across skill levels and technological familiarity.

\end{abstract}



\begin{teaserfigure}
    \includegraphics[width=\textwidth]{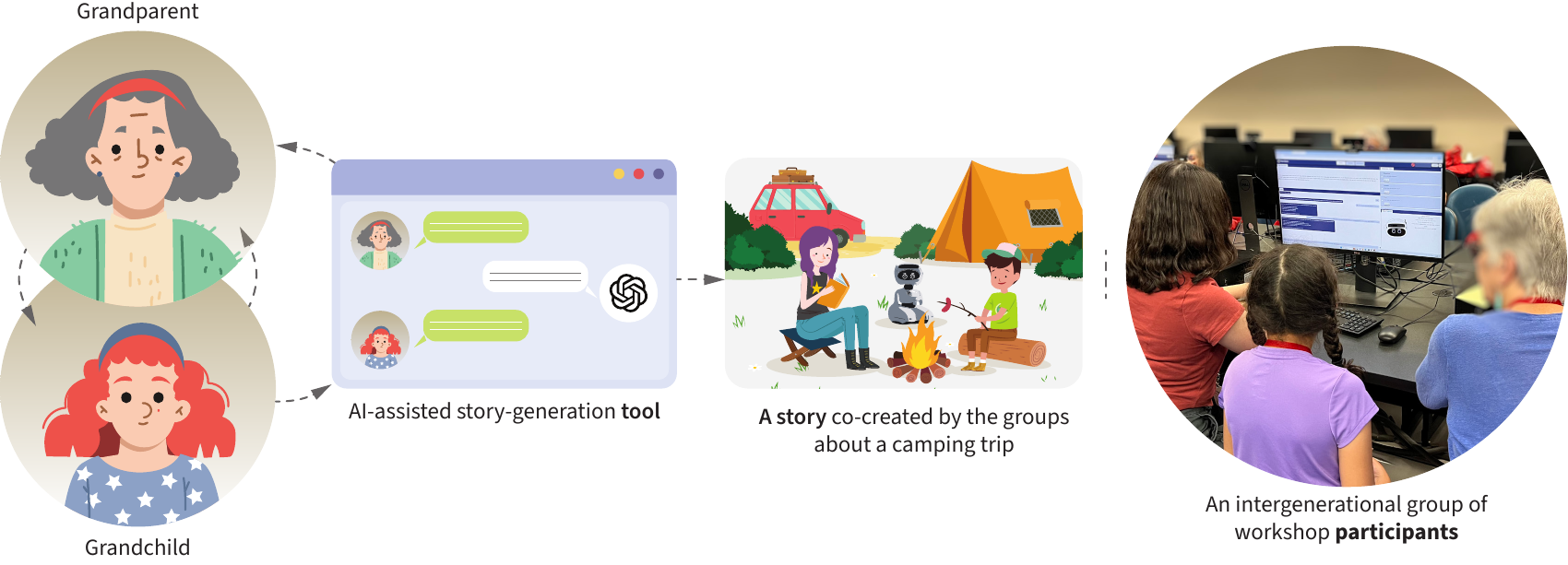}
    \caption{We explored AI's potential to support co-creation and strengthen communication across generations at a workshop. \textit{Left:} Intergenerational groups worked with an AI-assisted tool to co-create stories. \textit{Right:} A group of participants including a grandparent and two grandchildren from the workshop.}
    \Description{This figure illustrates the teaser of the paper. On the left, the figure depicts three entities involved in story creation: a grandparent, a grandchild, and an AI-assisted story generation tool, which serves as the interface. An example story is illustrated, connected from the interface, where two people are depicted camping with a robot in between. On the right side of the figure, a photo shows participants from the workshop interacting with the system through a computer screen, including one grandparent and two grandchildren.}
    \label{fig:teaser}
\end{teaserfigure}

\maketitle

\section{Introduction}

Intergenerational \camera{collaboration}{interaction} between grandparents and grandchildren can foster meaningful connections \cite{anderson2017translating} and strengthen family bonds \cite{cheng2024taking}. These connections can benefit younger generations socially and developmentally \cite{holmes2009intergenerational} and can contribute to self-efficacy, health, and well-being among older generations \cite{zhong2020intergenerational}. Although a positive and healthy intergenerational connection is universally desired, families face a number of barriers to establishing and maintaining such connections, including a mismatch between different generations in skill, language, interest, and availability \cite{strom2015assessment, barsukov2018barriers, kailen2020identity, 10.1145/3585088.3589358}. Grandparents may relish storytelling, cooking, or card games that their grandchildren may find boring, while grandchildren may enjoy video games, flying drones, and programming that their grandparents find too complicated or unfamiliar. Technology is a particularly notable barrier, as younger generations, raised in technology-rich environments, are more familiar with new technologies, whereas older generations may find catching up with the latest technologies challenging \cite{marzo2024bridging}. Researchers have called this phenomenon the ``digital divide,'' where significant differences in technology literacy and experience create barriers to effective communication and connection making \cite{volkom2014revisiting}.

Recent advances in artificial intelligence (AI), particularly large language models (LLMs), have the potential to overcome these barriers by substantially lowering the barrier to technology use through easy-to-use conversational interfaces that can handle a wide range of requests, questions, and tasks \cite{yang2024harnessing, kim2024exploring, kim2024understanding}. Applications such as ChatGPT \footnote{https://openai.com/chatgpt/} allow anyone with the ability to type or speak, regardless of age, background, and level of familiarity, to seek informational and creative help \cite{abdullah2022chatgpt}. Although users of these applications express concerns about their usability and potential to cause harm \cite{fischer2023generative, stahl2024ethics}, they are finding widespread societal adoption due to their potential benefits and ease of access. Recent research has reported the use of such tools for co-creation and co-exploration by families within conversations, shared tasks, and family entertainment \cite{10.1145/3613905.3650770}\camera{}{, pointing to an opportunity for AI to facilitate connection-making and to help overcome the digital divide. We focus on \textit{co-creation} as the process of bonding through shared creative activities that integrate the unique skills, experiences, and perspectives of both generations. Co-creation, facilitated by technology, can foster mutual learning and understanding, allowing grandparents and grandchildren to engage in storytelling, problem-solving, and technology use in a way that strengthens their relationship.  We use this term throughout the paper to emphasize the reciprocal and constructive nature of these interactions.} \revision{Therefore, we aim to answer the following research questions:}
\begin{enumerate}
    \item \revision{How do participants from different generations perceive AI's role within co-creation activities?}
    \item \revision{In what ways do AI-supported co-creation activities facilitate intergenerational communication and connection?}
\end{enumerate}

In this paper, we explore the potential of AI-assisted co-creation activities to establish intergenerational communication and connection and investigate how these activities shape intergenerational perceptions and perceptions of AI \camera{}{(Figure \ref{fig:teaser})}. For our exploration, we adopted an LLM-powered story-based robot programming tool and developed a semi-structured, workshop-based activity in which grandparent-grandchild pairs \revise{collaborated}{interacted} to co-create stories. The activity involved developing creative stories involving robotic and human characters---an activity grandparents might be particularly skilled and interested in---and translating these stories into programs that can control a social robot---an activity in which grandchildren might be particularly skilled and interested. The tool integrated an LLM assistant that helped pairs develop stories and translate them into robot programs.

In the context of a community outreach workshop, we asked 29 grandparent-grandchild pairs or groups (some groups involved more than one grandparent or grandchild) to develop stories and program a social robot based on these stories. We collected data on how the groups interacted, their development of the stories, their programming of the robot, their perceptions of the activities, and how the activity affected their perceptions of their grandparents or grandchildren. We report on findings regarding role delegation within groups, use patterns for AI within the co-creation activity, and intergenerational differences in AI use and acceptance. Our work makes the following contributions.

\begin{enumerate}
    \item \textit{Primary Contributions:} We gained empirical insights into how \revision{different generations perceive AI's role within co-creation activities and how these activities facilitate intergenerational communication and connection.
    }
    \revision{
    \item \textit{Secondary Contributions:} 
        \begin{enumerate}
         \item We developed a workshop-based activity to foster and study intergenerational interaction in a semi-structured environment, serving as a practical resource for future research.
         \item We developed design implications for AI-assisted technologies that target family communication and connection.
       \end{enumerate}
    }
\end{enumerate}



\section{Related Work}

\subsection{The ``Digital Divide''} \label{sec:background}

There is substantial evidence indicating that older adults generally find interacting with computers and learning new technologies more challenging than younger generations \cite{dickinson2007methods, broady2010comparison}. For instance, \citet{10.1145/3411764.3445702} showed that older adults often prefer one-on-one support when learning digital tools and may still prioritize non-digital tools over digital alternatives. However, there is growing interest among older adults to acquire basic technology skills, especially as they seek to stay connected with family and friends, and better organize their lives \cite{lobuono2019older}. This highlights the growing acknowledgment of technology's role in enhancing social connections. Moreover, technology plays a significant role in bridging generational gaps, enhancing intergenerational relationships by allowing older adults to engage in meaningful activities with younger generations \cite{freeman2020intergenerational}. Despite this potential, older adults tend to use digital devices and software less frequently and report lower confidence levels when using technology, further emphasizing the need for accessible, supportive digital tools \cite{kuek2020healthcare}.

\subsection{Intergenerational \camera{Collaboration}{Interaction through Shared Activities}}


\revision{
Engaging in \camera{collaborative and joint}{shared} activities creates a common ground for participants, fostering more effective communication and promoting positive social interactions.
}
Previous research has explored various \camera{collaborative}{joint} activities, such as \revision{storytelling \cite{boivin2023co, 10.1145/3687031}}, drawing \cite{alexenberg2004creating, knight2013small}, and \revision{playing games \cite{cui2023chatlaw, 10.1145/2470654.2466138, hausknecht2017blurring, derboven2012designing, 10.1145/1810543.1810550}}, as ways to foster meaningful connections between generations. 

\revision{
The relationship between grandparents and grandchildren is a particularly unique and enriching bond. Research highlights that interactions between people of different age groups promote social integration, foster intergenerational understanding, and enhance reciprocal learning \cite{taylor2005distance}. These interactions also provide opportunities to convey important social values, such as environmental stewardship and collective responsibility, further demonstrating their societal relevance. \citet{tsiloni2024psychosocial} examined the psychosocial effects of intergenerational learning among school-age children and older adults, demonstrating its positive impact on well-being. Similarly,
}
intergenerational activities have been shown to benefit both younger and older generations by reinforcing family bonds, enhancing reciprocal learning, increasing understanding between generations, and reducing social anxiety \revision{\cite{de2017benefits, axelrod2019intergenerational}}. 
However, generational gaps in technology usage, often referred to as the ``digital divide,'' can present significant challenges to successful \camera{collaboration}{interaction}.

For instance, \citet{10.1145/3544548.3581405} explored virtual reality as a medium for remote communication between grandparents and grandchildren. Their findings revealed a generational gap in preferences, as more than half of the grandchildren listed playing games as their favorite activity, yet many felt it was not suitable for interacting with their grandparents. This underscores differences in generational interests and technology skills, which can create barriers to intergenerational \camera{collaboration}{interaction}. 

Several studies have explored the use of tangible devices to bridge intergenerational gaps by facilitating storytelling and communication. For instance, \citet{li2019story} introduced Story-Me, a slot-machine-like device encouraging older adults in nursing homes to share life stories with their children. Similarly, \revision{\citet{wallbaum2018supporting} developed StoryBox, a device enabling grandparents and grandchildren to share photos, artifacts, and audio recordings, fostering playful expression and bridging technological divides.} \citet{10.1145/3294109.3300979} designed Slots-Memento, a tangible device for preserving and sharing intergenerational stories through photo displays and recorded narratives. These tools demonstrate how accessible, intuitive technologies can promote meaningful storytelling and strengthen intergenerational connections.

While these approaches have used technology to facilitate intergenerational interaction, our work shifts the focus to AI-assisted co-creation, offering a novel approach to \camera{collaborative}{joint} participation between generations. By leveraging AI's potential, we aim to create a more inclusive interaction across the digital divide.


\subsection{\camera{AI as a Collaboration and Co-Creation Tool}{AI as a Co-Creation and Interaction Tool}}

AI tools have been increasingly used to assist individuals in \camera{collaborative}{interactive} contexts, where AI systems work as partners to the human users in performing tasks \cite{10.1145/3334480.3381069, licklider1960man}. These tools often help in idea development \cite{choi2024creativeconnect, shin2023integrating, lloyd2022designing}, problem-solving \cite{xu2020ai}, human-AI communication design \cite{lee2024ai} and so on. 

With the proliferation of web-based information and the development of LLMs such as ChatGPT (GPT-3.5 and GPT 4 \cite{achiam2023gpt}) and Gemini \cite{team2023gemini}, recent LLM-powered tools are equipped with handling a variety of information and tasks. The domain in which the LLM-based tools are used varies from healthcare \cite{ghosh2024clipsyntel, han2024ascleai, goel2023llms} to law \cite{cui2023chatlaw, fei2023lawbench}, and also span from simple text summarization tasks \cite{lin2024rambler, zhang2024benchmarking}, to complex and creative tasks, such as brainstorming ideas \cite{shaer2024ai, memmert2023towards} or creating new content (\textit{i.e.,} co-creation) \cite{li2024we, tseng2024keyframer, zhang2024llm}. 

\revision{Due to AI-powered tools being more prevalent, some researchers have investigated users' perceptions of AI, including mental models \cite{johnson1983mental,norman2014some, bansal2019beyond, villareale2021understanding} and the extent of human control and trust in AI suggestions. \citet{gero2020mental} investigated people's mental models of AI in a cooperative word guessing game, revealing that people revise their mental models most when \camera{anomalies}{AI anomalies} persist. Collaborative settings, such as AI-assisted decision-making have also been extensively studied \cite{bansal2021does, ma2023should}. For instance, \citet{ma2023should} and \citet{bansal2021does} studied AI trust calibration through correctness likelihood (CL) and found out that users were able to accept AI suggestions more appropriately when required to gauge their own confidence levels and to think more critically about the AI's explanations. Another interpretation of trust towards AI was sought by \citet{shareef2021new}. Regarding AI trust particularly within the elderly population, they revealed elderly trust towards autonomous systems could be achieved if they were to be able to control the system and find a sense of belongingness. }


In our review of previous work, understanding the potential of AI tools within intergenerational groups is still at its nascent stage. \revision{Specifically, no studies have thoroughly examined the current trust and usage levels of AI applications across generations or how AI co-creation tools might bridge these generational gaps.}
Prior work primarily focuses on AI use cases involving individuals or groups of individuals (\textit{e.g.,} peers of similar age group) using the tools in various contexts. However, it often overlooks the broader impact these tools have on the individuals using them \revision{and their perceptions toward these tools.} Closest to our work are studies that observed AI tool usage within the education realm, where qualities such as self-efficacy, learning motivation, and growth mindset of the students were examined \revision{through} using the tools \cite{kim2024exploring, meyer2024using, kumar2023impact}. Our work differentiates itself from prior research in that we (1) focus on a specific group of cohorts consisting of grandparents and grandchildren and (2) investigate the usage patterns and the effects of AI tools on bridging generational gaps. Particularly, our work adds to the body of work discussing the potential and usefulness of AI tools in co-creation. To this end, our work aims to explore opportunities that lie within AI-supported systems \revision{to facilitate} \camera{collaboration}{interaction} between the older and younger generations. 






\section{Methodology}
\revise{The authors note that two submissions have been derived from the same community outreach event. Although both submissions adhere to the same study protocol, they focus on distinct topics, addressing different research questions. This work focuses on co-creation activities with intergenerational groups, specifically grandparents and grandchildren, whereas the other submission focuses on end-user programming for novices. }{We conducted a user study in a workshop setting with 29 grandparent-grandchildren groups to explore intergenerational interactions. Each group was tasked to co-create a story with AI assistance through activities, such as storyboarding, writing, and programming, in the context of a workshop about social robotics. The workshop took place in a suburban college town community.}
This user study procedure was approved by the authors' Institutional Review Board.

\subsection{Participants}
All participants were informed about the study's goals and given the option to volunteer for the workshop.
The outreach program was advertised to grandchildren between the ages of 7--14 \revise{}{($M = 10.64, STD = 2.13$)} and their grandparent(s) \revise{}{($M = 74.07, STD = 7.09$; 2 participants did not report their age)}.
\revise{}{A total of 67 participants across 29 groups were recruited, comprising 31 grandparents (12 male, 19 female) and 36 grandchildren (24 male, 12 female). Groups} were composed of grandchild(ren) and grandparent(s), with varying numbers in each category (\textit{e.g.}, one grandparent and one grandchild, one grandparent and two grandchildren, etc.).\revise{}{ All participants were required to be fluent in English. There were no requirements to have any technological familiarity with programming or robotics prior to the workshop. All grandparents were either alumni of or related to an alumni of \camera{a college town located in a suburban community}{the University of Wisconsin--Madison}.} All participants were recruited through an alumni organization.
Beyond the learning outcomes of the outreach event, no additional compensation was provided to the participants.

\begin{figure*}[t]
    \centering
    \includegraphics[width=\textwidth]{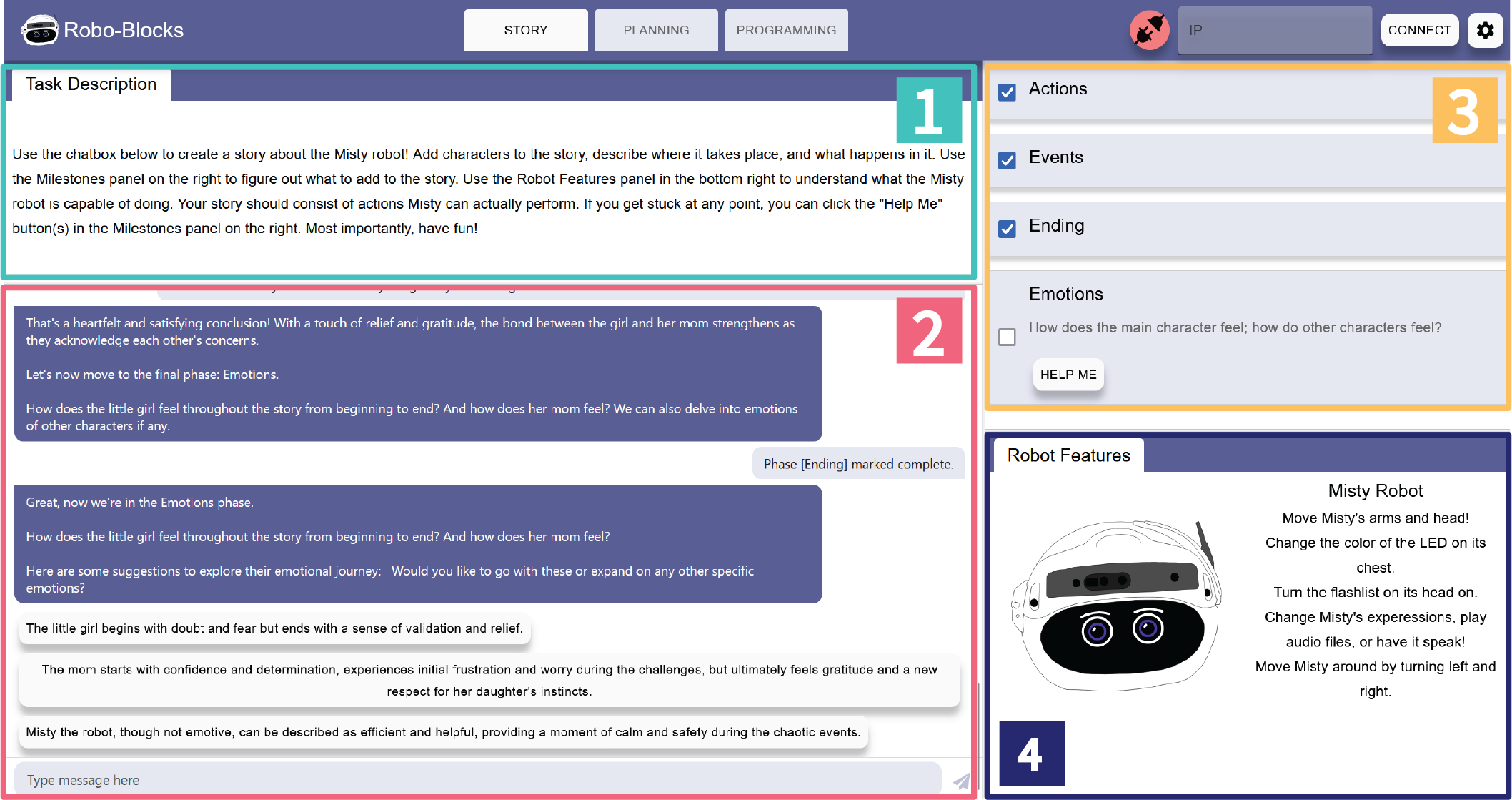}
    \caption{Overview of \oursys. (1) description of the task; (2) chat interface; (3) milestones with a ``Help me`` button; (4) description of the robot platform.}
    \Description{The figure shows the designed user interface. At the top left, there is the task description area, explaining the activity and interface features. The bottom left features the chat interface for user interaction. The top right displays a checklist that outlines milestones in the story creation process. Finally, the bottom right section shows an image of a robot along with a text description of its capabilities.}
    \label{fig:interface}
\end{figure*}

\subsection{\oursys: LLM-powered Story-based Robot Programming Tool} \label{sec:system}

We describe \oursys \camera{}{(Figure \ref{fig:interface})}, an LLM-powered co-creation tool for writing and programming. 
Below, we explain the features of ~\oursys that are relevant within the scope of this work:

\textbf{Task Description:}
\oursys provides a detailed description of the story creation task along with explanations of its features, which can be accessed at any time for reference to provide and remind users of the task.

\textbf{Chat:}
\oursys supports a chat-based interaction for the story creation task.
Powered by GPT-4o, \footnote{https://openai.com/index/hello-gpt-4o/}, users can \revise{collaborate}{interact} with an LLM agent through the chat interface to create a story. 
\revise{}{To begin the interaction, the user begins by typing in details of their story and then sends a message to the LLM agent. Users can send a message to clarify the details of the task, ask for suggestions on story creation, and support for editing and organizing story ideas. The LLM agent then responds by asking further guiding questions and the user continues to fill out details of the story in this manner.}

\textbf{Milestones:}
\oursys provides users with milestones grounded in the Self-Regulated Strategy Development (SRSD) ~\cite{Mason2002_StoryMilestones} to provide structure to the writing process and support story creation.
The milestone provides guidance by identifying seven elements for the user to consider: \textit{characters, location, time, actions, events, ending, and emotion.}
The interface allows users to toggle each element to indicate completion. 
\revise{}{After marking an element complete, the LLM agent provides guiding questions for the users to consider for the next element.}
Below each element, there is a \textit{``Help me''} button that enables users to request suggestions from the LLM agent. Upon clicking the button, users receive clickable suggestions \revise{}{from the LLM agent }on how to fill in details for the selected element. 

\textbf{Robot Description:}
\oursys displays the robot's capabilities and design, allowing users to integrate the robot as a main character in their story. As part of the story creation process, participants write executable actions that will be programmed and deployed during Day 2 activities.

\subsection{Community Outreach Event}
The authors coordinated with a community partner to organize a community outreach event to teach social robotics and programming for families, focusing on grandparents and grandchildren working together.\footnote{Materials for the curriculum, including lecture slides, prompts, and workshop schedule can be found at \url{https://osf.io/t76ys/?view_only=ea243a1442d548e2908b3a1791c67c90}.}
The outreach event was a two-day session offered once a week, held over three weeks for a total of (three sessions $\times$ two days) six days. 
Each week featured a new group of participants and was facilitated by one instructor and two-to-three helpers who provided assistance to the group during activities. 

The primary objectives of the event were to:
\begin{enumerate}
    \item Facilitate intergenerational learning and interaction through AI and robotics activities.
    \item Study how different generations \revise{collaborate}{interact} in a semi-structured co-creative environment.
    \item Provide participants, especially older adults, opportunities to engage with emerging technologies.
\end{enumerate}
The curriculum was structured to balance both structured learning (\textit{e.g.}, lectures) and open-ended co-creation (\textit{e.g.}, AI-assisted story creation and robot programming). This workshop-based activity provided an opportunity to systematically study intergenerational \camera{collaboration}{interaction} within a semi-structured environment. By combining structured learning with open-ended exploration, the workshop facilitated meaningful interaction and provided insights into how different generations \revise{collaborate on}{interact during} co-creation tasks.




\subsubsection{Day 1: Story Creation\camera{and Intergenerational Collaboration}{}}
The primary focus of Day 1 was to foster intergenerational learning and \camera{collaboration}{shared creativity} through a story creation activity. The session began with an introductory lecture on social robotics, covering key topics such as the basic principles of robots, social behaviors of robots, and human-robot interaction. The lecture was designed to create a shared understanding between generations, as prior research has shown that collaborative learning experiences can enhance both engagement and outcomes in intergenerational settings \cite{aemmi2017intergenerational, doi:10.1080/15350770.2020.1817830, corrigan2013intergenerational, hammaren2022scoping}. 

To initiate interaction and shared storytelling, participants were tasked with a storyboard activity. In this exercise, groups reflected on family memories and brainstormed ideas for stories in which robots could play a role. This activity was designed to build rapport within groups while setting the stage for creative co-creation in a shared narrative context.

Participants were then given a 20-minute lab tour, and were invited to interact with social robot Pepper. This hands-on experience served to demystify technology for older adults, helping them feel more comfortable with robots in a casual setting.

Following this, a tutorial on \oursys was provided, demonstrating the interface's features and introducing LLMs concept and ChatGPT. The AI-assisted story creation task concluded Day 1, where groups used \oursys to co-create their own stories involving a Misty robot as one of the main character. 

\subsubsection{Day 2: Programming and Story Deployment}
On Day 2, the focus shifted from story creation to programming, continuing the semi-structured format by having participants program a Misty robot to act out the story they had created on Day 1. The session began with a tutorial on \oursys, where instructors demonstrated how to use the programming interface. Participants were guided through examples to ensure that all groups could successfully follow the instructions.

After the tutorial, participants worked in groups to program the robot. Once robot programming was finished, each group deployed their program to a physical robot. This phase combined story creation with technical engagement, allowing participants to see the tangible outcomes of their \camera{collaboration}{co-creation}.

To encourage further exploration and engagement, participants were provided with slides that outlined advanced tasks and design choices for programming the robots. This open-ended exploration phase created further opportunities to study intergenerational \camera{collaboration}{co-creation} in a semi-structured environment.


\revise{All participants were informed about the study's goals and given the option to volunteer for the workshop.
The outreach program was advertised to grandchildren between the ages of 7--14 and their grandparent(s).
We recruited 29 groups, which were composed of grandchild(ren) and grandparent(s), with varying numbers in each category (e.g., one grandparent and one grandchild, one grandparent and two grandchildren, etc.).
All participants were recruited through an alumni organization.
Beyond the learning outcomes of the outreach event, no additional compensation was provided to the participants.}{}

\subsection{Experimental Setup}
All groups regardless of their decision to participate were provided a computer station that consisted of a desktop computer, one keyboard, and one mouse. 
For groups that opted to participate in the user study, an additional web camera was mounted on top of the computer display to record the interaction of the groups as they worked with one another. Additionally, participants had their computer screens recorded to see how they used the system.
To avoid recording groups that chose not to participate, groups of study participants were positioned in the same area when possible.

\subsection{Procedure}
All consent for participating grandchildren was received in advance from their parents and oral assent was confirmed
 prior to recording their activities. 
 Consent from the grandparents was received on the first day of the workshop.
After confirming consent from the participants, the experimenter introduced the purpose of the user study and activities.
All participants, regardless of participant in the user study were provided the same tasks and resources.
The workshop activities were distributed across two days. 
Here, we refer to all attendees of the workshop, regardless of their participation in the user study, as groups.

\textbf{Day 1:} The instructor begins the workshop by teaching a 40-minute introductory course about social robots, that includes examples of social robots, a brief discussion of robot design, the uncanny valley, and a discussion of social cues and behavior.
The groups are then given 40 minutes to discuss and create a storyboard about a family memory or experience, emphasizing any social cues in the storyboard. 
Following this activity, the instructor and helpers present an interactive demonstration with an LLM-powered robot for 20 minutes. 
Then, during a 15-minute break, the instructors and helpers confirm oral assent from the grandchildren and written consent from the grandparents.
Groups that opted to join the study were then guided to complete the technology familiarity questionnaire (Section \ref{sec:sec_pre-survey}).
The instructor leads a 20-minute tutorial on how to use the \oursys interface, focusing on the chat interface for story creation.
The groups are then given 30 minutes to create a story based on the previously made storyboard using the \oursys interface (refer to Section \ref{sec:system}).
Finally, Day 1 concludes with a 10-minute wrap-up session where groups are encouraged to share their story aloud.

\textbf{Day 2:} 
The instructor commences Day 2 with a 30-minute tutorial on how to use the programming interface of \oursys (refer to Section \ref{sec:system}).
The groups are given 30 minutes to engage in robot programming.
\revise{Participant groups are given the post-task survey (Section \ref{sec:post-task-survey}).}{}
Following the programming session, the instructor and helpers set up the robot programs to be deployed on the Misty robot. 
After the deployment, the instructor and helpers conduct individual group interviews (Section \ref{sec:group_interview}).  






\subsection{Data Collection}

For groups that consented, the story writing and robot programming exercises were audio, video, and screen-recorded.
All group interviews were conducted as families and audio-recorded for later analysis. Data collection was done using \revise{five}{four} methods: (1) technology familiarity questionnaire; (2) video and screen recording; (3) activity outcomes; \revise{(4) post-task survey; and (5)}{and (4)} group interviews. We discuss each method in detail below. 

\subsubsection{Technology Familiarity Questionnaire} \label{sec:sec_pre-survey}
We distributed a questionnaire (refer to Appendix \ref{sec:pre-survey}) to capture familiarity and experience with ChatGPT and with programming. 

\subsubsection{Video and Screen Recording}
During the story creation and robot programming activity, we recorded both a video facing the groups to observe their interaction with one another and usage of keyboard and mouse, and a screen recording to confirm their interactions with \oursys.

\subsubsection{Activity Outcomes}
We logged the history of the chat communication with the LLM agent during the story creation activity \revise{and the final program from the programming activity.}{ to verify story creation and milestone usage.}


\subsubsection{Group Interviews} \label{sec:group_interview}
Over the course of three sessions, four of the authors and two external individuals experienced with conducting interviews, hereinafter referred to as interviewers, conducted semi-structured interviews with individual groups of participants about their experience using \oursys.
The interview questions are available in Appendix \ref{sec:interview}.
Due to a large number of groups and time constraints, members of the same family took part in a group interview together (6--19 minutes).
All interviewers followed a semi-structured interview script to ensure consistency across interviews.


\subsection{Analysis}
We analyzed all collected data\revise{, except the post-task survey as the results and findings are outside the scope of this work}{}, which includes 26 interviews from 29 groups. In one instance, two groups from the same family were interviewed together, one group declined to participate, and one interview was missed.
We then conducted a thematic analysis~\cite{McDonald19, clarke2014thematic} on the interview data combining an inductive (data-driven) approach and a deductive (question-driven) approach. 
The first two authors independently approached the interview data with an inductive approach on a subset of three interviews to derive an initial draft of a codebook \cite{decuir2011developing}.
Examples of codes in common between coders included ``role while using \oursys,'' ``impression of AI,'' and ``impact of collaboration with grandparent/grandchild.''
The coders divided the remaining interviews and used the codebook to drive a deductive analysis of the remainder of the interviews.
The coders then revised the other's codes and added new codes that helped characterize the interaction between grandparents and grandchildren and their perception of AI support.
\revise{We also recruited two external coders to do the initial review video and screen recordings to confirm the groups' usage of \oursys, engagement with AI, and the role of each group member during the activity.}{With the video dataset, we recruited two external coders to do the initial content analysis to confirm the groups' usage of $\oursys$, engagement with AI and milestones, and the role of each member during the activity. The first two authors later conducted a follow-up content analysis, focusing on the interaction types between participants and between the participant(s) and AI.}
For quantitative data collected from the pre-survey, a chi-square test of independence was conducted to examine the relationship between generation (Grandparents vs. Grandchildren), ChatGPT usage (Yes vs. No), and programming experience (Yes vs. No). 


We acknowledge that our findings of the opinions and experiences of the grandparents and grandchildren participants in this study do not represent absolute truths on how other grandparents and grandchildren may use AI-supported interfaces for co-creation.
Rather, our dataset captures various ways in which grandparents and grandchildren may perceive and respond to support from one another and AI. 

\section{Findings} \label{sec:findings}

\begin{figure*}[t]
    \centering
    \includegraphics[width=\textwidth]{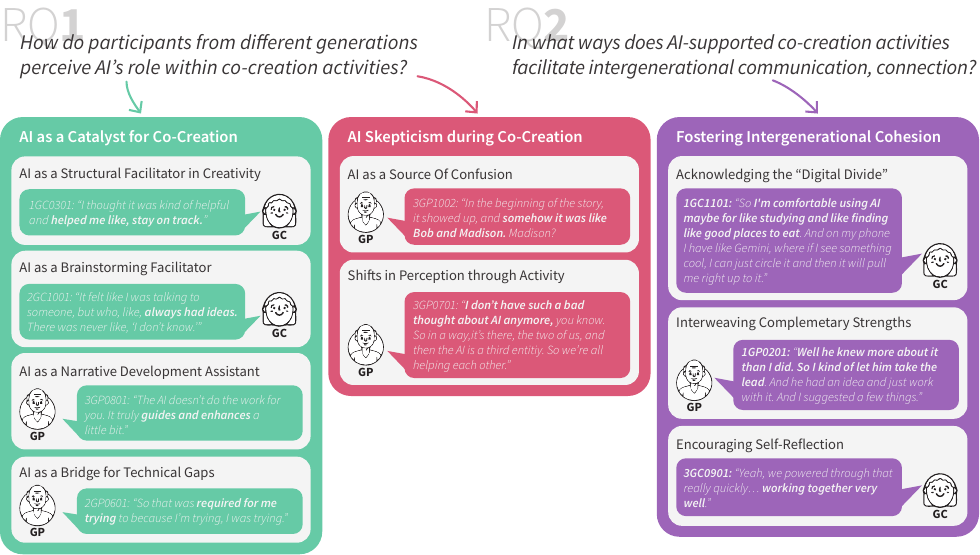}
    \caption{\revision{
    Summary of Qualitative Findings: AI was perceived as a catalyst by structuring the process, supporting brainstorming, enhancing storytelling, and bridging technological gaps, fostering mutual appreciation between generations. Also, participants exhibited mixed perceptions of AI during co-creation activities across generations. Finally, AI-assisted co-creation activity fostered complementary strengths between generations by creating a shared environment, allowing both generations to contribute their unique strength and encourage self-reflection. Background shapes connect themes related to the research questions. GP denotes grandparent, and GC denotes grandchild.
    }}
    \Description{A graphical summary of the qualitative analysis, depicted with three themes. The left theme explains AI's perceived role within groups. The middle theme talks mixed perceptions of AI across generations. The right theme shows quotes relevant to fostering intergenerational cohesion through AI-assisted story co-creation. Research question 1 points to the left and middle theme while research question 2 points to the right theme.}
    \label{fig:findings}
\end{figure*}

\revision{We present findings derived from both quantitative and qualitative data analyses. Results from the video analysis are summarized in Tables $\ref{tab:interactions_human}$. Table $\ref{tab:interactions_human}$ outlines key interaction scenarios observed between grandparents and grandchildren during the task. 
From these analyses, three main themes emerged, each offering a deeper understanding of how AI-assisted co-creation activities foster intergenerational communication, enhance connection, and shape perceptions of AI. To maintain participant anonymity, grandparents are coded as ``GP,'' and grandchildren are coded as ``GC.''}


\subsection{\revision{AI as a Catalyst for Co-Creation}}
\revision{Participants perceived AI as playing four key roles within co-creation activities.} First, AI provided a structured framework, helping participants stay focused and organized during story creation. Second, AI enabled collaborative brainstorming by offering prompts that sparked discussions and encouraged idea-sharing between grandparents and grandchildren. Third, AI enhanced storytelling by offering suggestions and refining inputs, improving the overall quality of the narratives. Finally, AI's navigational support bridged technological gaps, allowing both generations to contribute meaningfully regardless of their digital literacy. \revision{These roles highlighted how participants viewed AI as a mediator that leveraged the unique }strengths of each generation, fostering engagement and strengthening the bond between grandparents and grandchildren.

\subsubsection{\revision{AI as a Structural Facilitator in Creativity}}
Participants perceived AI as a structural facilitator as AI provided a structured framework for story creation, using pre-designed milestones provided through initial prompts. These milestones allowed the AI to track progress and offer guidance that aligned with the intended flow of the activity. This structure helped participants to stay focused on specific parts of the story, preventing them from feeling overwhelmed. 
One grandchild reflected on this structured approach, \textit{1GC0301: ``I thought it was kind of helpful and helped me stay on track and know what I'm supposed to do for this story.''}
Grandparents also found the structure beneficial as a tool for ensuring important story elements were not overlooked. As one grandparent noted, \textit{1GP0901: ``It was a good reminder that you're not skipping over who the main character is, where it's happening, what the feelings are...''} Another grandchild highlighted how the milestones gave direction throughout the process, sharing, \textit{3GC0901: ``We had... milestones tell us what directions we should be using.''}
However, not all participants found the structure equally useful. One grandparent, an experienced writer, found the milestone system rigid, commenting, \textit{2GP0701: ``If somebody was not a writer, who hadn't written lots and lots of stories, then that step-by-step process, I think would be useful. But to me, [the milestones were] so plottingly slow because I would automatically go through what those [milestones] were asking you to do.''}

\subsubsection{\revision{AI as a Brainstorming Facilitator}}
Participants perceived AI as a collaborator that facilitates brainstorming by asking follow-up questions that sparked discussions between grandparents and grandchildren. This encouraged both grandparents and grandchildren to share ideas and contribute diverse perspectives. One grandchild reflected on how AI enhanced the flow of idea generation, \textit{2GC1001: ``It felt like I was talking to someone... who always had ideas. There was never [an], `I don't know.'.''}
AI's prompts not only guided the storytelling process but also motivated both generations to dive deeper into the story details. Another grandchild shared, \textit{3GC0701: ``They helped us interact because it would remind us that we needed more evidence, like, what is this person doing? Or who is this person? What age is this person?''} 
This interaction led to smoother brainstorming, making it easier for both generations to build on each other’s ideas seamlessly and an enjoyable creative experience for both parties. One grandparent expressed enjoyment in brainstorming ideas with the assistance of AI, noting,
\textit{2GP0801: ``That was fun (grandchild says this at the same time). We went silly in a hurry.''}
AI's prompts not only encouraged brainstorming but also helped clarify story elements that could be confusing for readers of the created story, ensuring the narrative was coherent and well-structured. As another grandparent remarked, \textit{2GP0701: ``It was useful for helping us clarify, and avoid misunderstandings that our reader might have.''}

\subsubsection{\revision{AI as a Narrative Development Assistant}}
Participants perceived AI playing a pivotal role in refining the co-created stories by offering suggestions and filling narrative gaps. One grandparent highlighted how AI contributed creative suggestions to enhance the story, stating, \textit{3GP0201: ``I think it was helpful to be able to get the hints to flesh out the story because ours wasn't entirely, an experience that we had together.''}
When participants got stuck, AI provided multiple solutions to help move the process forward. As one grandchild noted, \textit{2GC0101: ``That's[AI] really helpful because we got we got stuck trying to find a solution... it was very helpful [and gave] us multiple solutions to our problems.''} 
In some cases, AI took basic outlines and expanded them into richer narratives, as one grandparent explained, \textit{3GP0901: ``So I think that what we had typed in was more or less of an outline. And then when the AI took the outline, and made it into a story, it fleshed it out, and he[AI] has a lot of extra content, and made it a better story than what we were doing with it.''}
Participants also appreciated AI's flexibility and responsiveness, especially when they made errors or wanted to revise their work. One grandchild shared, \textit{2GC1002: ``I had trouble. I accidentally would press the wrong thing and it[AI] would do it[create story] but then I really said I wasn't finished with that part and it would just put it back.''}
Another grandchild remarked on how quickly AI could adjust the narrative, \textit{2GC1001: ``Now write this all but from Misty's perspective, story, (2GP1001 notes it's interesting) and then it just changed a little thing for me. And then to make it shorter, just a matter of seconds.''}
Participants emphasized that while AI offered helpful suggestions, it never took over the process. Instead, it enhanced their inputs and allowed them to remain in control. As one grandparent shared, \textit{3GP0801: ``The AI doesn't do the work for you. It truly guides and enhances a little bit.''} Another grandparent echoed this, \textit{1GP0201: ``He put his ideas down, and it was rearranged into a narrative that was quite nice. It's still what his ideas were, you know, I liked how that turned out.''}

\subsubsection{\revision{AI as a Bridge for Technological Gaps}}
Participants perceived AI as a navigator in technology. In AI-assisted co-creation, the ``Help me'' feature played a crucial role in supporting participants who were less familiar with digital technology, ensuring both generations could contribute meaningfully to the storytelling process. This AI-driven function provided intuitive guidance, offering suggestions and direction when participants encountered challenges. By simplifying the interface and allowing participants to request assistance from the AI system at any stage, the feature helped bridge technical gaps, enabling smoother \camera{collaboration}{co-creation}. One grandparent emphasized the value of this support, sharing, \textit{2GP0601: ``when we clicked on help me it gave us suggestions, and it's based on an AI. I did all the time. So I thought that helped me was very helpful because I had no idea what to do next. So that was that was required for me trying to because I'm trying, I was trying.''} With the AI providing navigational support, participants who might have otherwise struggled with the technical aspects of the activity were able to engage in the creative process without feeling overwhelmed.

\begin{figure*}[h]
\centering
  \includegraphics[width=\textwidth]{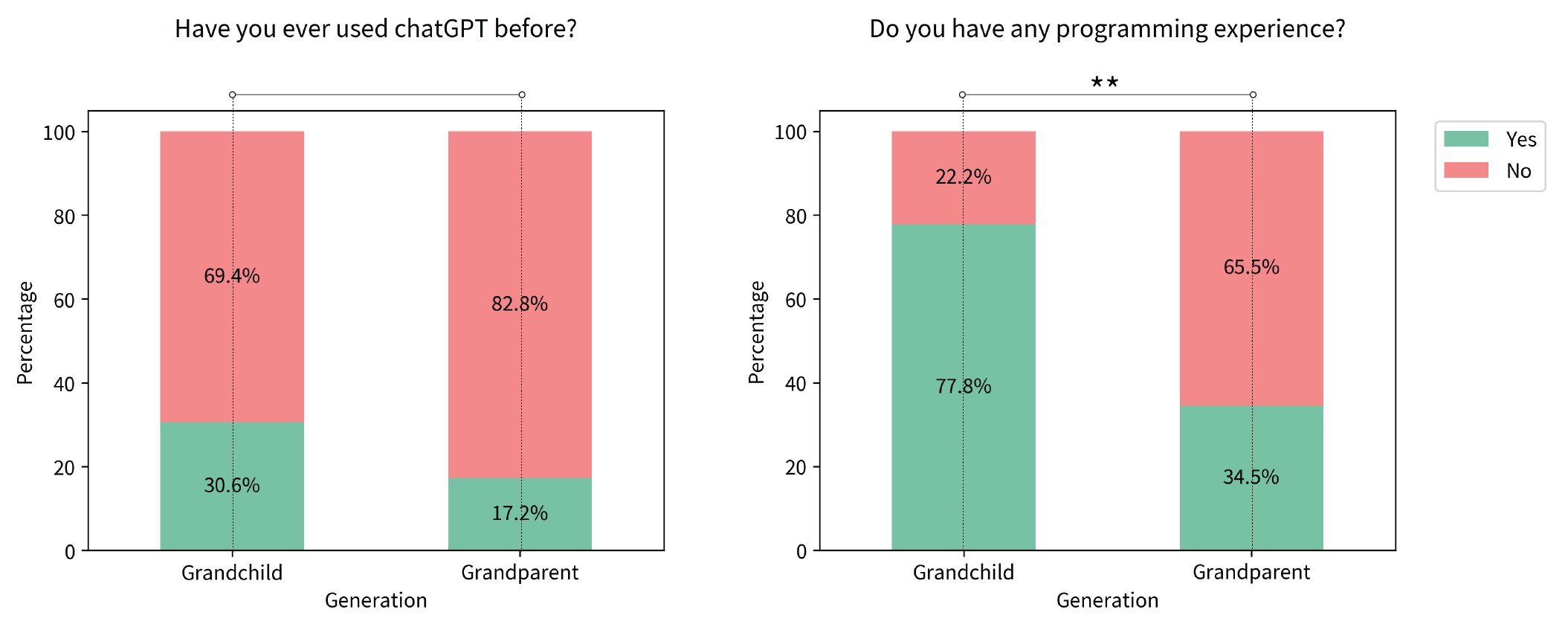}
  \caption{Comparison of ChatGPT usage and programming experience across generations. Horizontal lines indicate statistically significant differences based on the Chi-Square test ($p < .05^{\ast}$, $p < .01^{\ast\ast}$, $p < .001^{\ast\ast\ast}$).}
  \label{fig:pre-survey}
  \Description{The left figure displays the distribution of 'yes' and 'no' responses regarding ChatGPT usage across different generations. The right figure shows the distribution of 'yes' and 'no' responses related to programming experience across generations. There is no significant difference in ChatGPT usage between generations, with both showing low levels of usage. However, there is a noticeable difference in programming experience, with grandchildren generally having more experience compared to grandparents.}
\end{figure*}

\subsection{\revision{AI Skepticism during Co-Creation}}
\revision{
Participants exhibited mixed perceptions of AI during co-creation activities. When AI generated a response that is unfaithful to the source input, it caused confusion among the participants. Conversely, when AI responses aligned with expectations, participants found the experience enjoyable. While grandchildren generally viewed AI as a valuable creative partner, some grandparents expressed discomfort and fear. However, hands-on engagement led many grandparents to develop a more positive and nuanced understanding of AI, demonstrating that collaborative experiences can help bridge generational gaps in AI acceptance.
}

\subsubsection{\revision{AI as a Source of Confusion}}

\revision{Participants encountered confusion when AI hallucinated, such as including a character in the story not added by participants. For instance, one grandparent noted, \textit{3GP1002: ``In the beginning of the story, it showed up, and somehow it was like Bob and Madison. Madison? So it was something either we said or did or typed in inadvertently. And it got put into the story character.''}} \revision{Although these characters were not introduced by the participants, they attributed the error to their own input rather than recognizing it as an AI-generated mistake.}

\revision{However, when AI responses was aligned with their expectations, the experience was enjoyable. For instance, one grandparent noted,}
\textit{1GP0901: ``It seemed like it was already something that as long as you're putting in the sequence, and the way the computer can understand it, and the way you think the computers can understand it, then it'll work how you're expecting. So... I think it was just for me, it was a very enjoyable experience.''} 

\subsubsection{\revision{Shifts in Perception through Activity}}
\revision{
After the activity, 13 grandchildren shared their positive experiences with AI. One grandchild shared, \textit{2GC0501: ``I really like AI. It’s really helpful. I think that it helps write a story really well.''} Another remarked, \textit{2GC1001: ``I thought it was helpful. It felt like I was like talking to a person''} A third grandchild added,
\textit{1GC0302: ``Very helpful, very helpful when it's giving you ideas. It gives you like four or three ideas.''} These comments suggests that younger participants viewed AI as a valuable creative partner that enhanced the co-creation process by providing ideas and solutions.
}

Four grandparents found AI to be a baffling and unfamiliar concept. One grandparent expressed discomfort with AI-generated stories, stating, \textit{1GP0901: ``It's hard for me to think about using AI for generating stories because I've realized that in my mind, I still put a lot of meaning and value into somebody doing the work on their own. But I also think that part of that is just due to my generational experience.''} Another grandparent expressed skepticism, questioning the reliability of AI, \textit{1GP1001: ``Where does it get the information? I mean, where does it all that come from? Who's supplying it? I have an inclination not to trust it.''} Concerns about AI were also tied to broader societal fears. One grandparent admitted, \textit{1GP1101: ``We're scared of it. You know, as my friends as teachers and that we are fearful about it being used in our society. I would never use it in my life.''} However, throughout the co-creation activity, grandparents reported changes in their perception of AI. Eight grandparents expressed that their views of AI changed as a result of the co-creation activity. One grandparent, who had limited prior knowledge of AI, found the experience to be enlightening.
\textit{1GP1101: ``It's been eye opening for me, you know, to get into an area that I'm not involved in or ever have been involved in. And so I can be maybe a little bit more intelligent when people are talking about, some of this reading about it, or it's on the news or something. It's given me good background information.''} Another grandparent described a significant shift in their attitude toward AI from skepticism to seeing AI as a helpful third entity. \textit{3GP0701: ``I don't have such a bad thought about AI anymore. you know. So in a way, it's there, the two of us, and then the AI is a third entity. So we're all helping each other.''}

\revision{
This shift indicates that generational differences toward perceptions of AI exist but could be demystified through positive, hands-on co-creation experiences.
}

\begin{table*}[h]
    \caption{\revision{Interaction scenarios observed between grandparents and grandchildren in video recordings of 29 groups during the user study.}}
    \label{tab:interactions_human}
    \centering
    \renewcommand{\arraystretch}{1.2}
    \small
    \begin{tabular}{p{0.2\linewidth}p{0.5\linewidth}p{0.2\linewidth}}
    \toprule
        \textbf{Interaction Category} & \textbf{Observed Behavior} & \textbf{Interaction Direction}  \\ \toprule
        Collaborative Ideation & Soliciting ideas for the story & GC $\Longleftrightarrow$ GC or GP  \\ 
        Praise/Compliment & Praising or complimenting the other person’s idea or skill & GC $\Longleftrightarrow$ GP \\
        Encouragement & Encouraging the other to contribute or share ideas & GP $\Longrightarrow$ GC \\ 
        Shared Laughter & Laughing together while discussing ideas & GC $\Longleftrightarrow$ GP \\ 
        Language Assistance & Requesting help with spelling or grammar & GC $\Longrightarrow$ GP \\ 
        Clarification Requests & Clarifying AI responses or intentions & GP $\Longleftrightarrow$ GC \\ 
        Technical Assistance & Seeking help with typing tasks or using the computer mouse & GC $\Longrightarrow$ GC or GP \\
    \bottomrule
    \end{tabular}
    \Description{Table describes interaction scenarios observed between grandparents and grandchildren in video recordings of 29 groups during the user
study. It includes three columns: Interaction Category (types of interactions observed), Observed Behavior (specific behaviors exhibited during the interaction), and Interaction Direction (who initiated the interaction and whether it was mutual or unidirectional). Existing categories are collaborative ideation, praise/compliment, encouragement , shared laughter, language assistance, clarification requests, and technical assistance. Interaction directions indicate reciprocity with double-sided arrows or one-sided direction with one-sided arrows.}
\end{table*}

\subsection{\revision{Fostering Intergenerational Cohesion Through AI-Assisted Story Co-Creation}}
\revision{
AI-assisted co-creation activity fostered complementary strengths between generations. Grandchildren's digital fluency facilitated interaction with AI, while grandparents contributed narrative depth and personal insights. This synergy enabled meaningful involvement and mutual appreciation. Feedback from participants suggested that grandchildren came to recognize the value of their grandparents' storytelling ability, while grandparents reported finding technology more approachable with their grandchildren's guidance. These observations point to a reduction in intergenerational tension and a development of a more harmonious \camera{collaboration}{working partnership}. By providing a structured, AI-assisted environment, the activity highlighted complementary skills and encouraged a positive and collaborative intergenerational dynamic.
}

\subsubsection{Acknowledging Generational Differences in Technological Familiarity}

Our pre-survey data revealed differences in programming experience between grandparents and grandchildren, but no significant difference in ChatGPT usage (Figure~\ref{fig:pre-survey}). While grandchildren had generally higher exposure to programming (Chi-square test: ${\chi}^2(1, N = 65) = 10.68$, $p = 0.001$), ChatGPT usage was reported similarly across generations. Differences in technical familiarity influenced the roles participants assumed during the co-creation activity and shaped their perspectives on AI’s role. 

Qualitative feedback from interviews provided further insights into these differences, \revise{}{from familiarity to computer hardware to AI technology}. 
Eleven grandchildren noted early exposure to digital technologies in educational settings 
While grandchildren were generally comfortable with touch-based interfaces on tablets or phones, some found traditional keyboards and mice less intuitive. 
In contrast, grandparents exhibited varying levels of technological familiarity, often acknowledging the grandchildren’s greater digital competence. As one grandparent noted, \textit{1GP0101: ``He has more experience. I never had that kind of stuff. He's got more experience with it.''} 
\textit{1GP0201: ``And it's that age now, you know, and I did not grow up with computers and robots. So it's pretty foreign.''} 

Regarding AI specifically, younger participants expressed confidence and enthusiasm. For example, one grandchild indicated comfort using AI daily for studying or finding recommendations on a mobile device equipped with an AI assistant \textit{1GC1101: ``So I'm comfortable using AI maybe for like studying and like finding like good places to eat. And on my phone I have like Gemini, where if I see something cool, I can just circle it and then it will pull me right up to it.''} 
\revision{Another grandchild mentioned exposure to AI through a parent's job, noting familiarity with AI-generated images, \textit{1GC0101: ``I have from my dad's job to work in AI. So I've seen him generate photos, but I've used it.''}} However, some grandparents who had encountered the concept of AI decades ago were impressed by its current capabilities, \textit{2GP0101: ``I went to (university name) in the 60s. And the word was out at that time, it was artificial intelligence. But without the computer power that exists since that time, it couldn't do a lot more.''}
\revision{
These findings highlight a complex interplay of experience, confidence, and perceptions from different generations, influencing roles participants assumed during the activity. While younger participants often led the interaction with novel interfaces and AI tools, grandparents contributed valuable insights, enabling both generations to collaboratively shape their co-creation experience.
}

\subsubsection{\revision{Interweaving Complementary Strengths}}

As participants engaged in co-creation, they fluidly negotiated roles, responding to each other's strengths and limitations.  Grandchildren, generally at ease with digital navigation, primarily managed technical tasks such as interacting with the AI and entering text. Grandparents, in turn, contributed creative direction and narrative guidance. One grandparent remarked, \textit{3GP0401: ``He was way ahead of me in just about every step of the process.''}, while another added, \textit{1GP0201 ``Well he knew more about it than I did. So I kind of let him take the lead. And he had an idea and just work with it. And I suggested a few things.''} 
\revision{highlighting how grandparents enriched the process with narrative input. However, these roles were not static. In some instances, grandchildren struggled with certain interactions—such as typing on a keyboard or using a mouse, having been more familiar with touchscreen devices which prompted grandparents to step in. This demonstrates that generational roles were not rigidly defined but dynamically adjusted for mutual benefit.}

\revision{
Table~\ref{tab:interactions_human} shows various scenarios that emerged, further enriching this \camera{collaboration}{interaction}. \textbf{Collaborative Ideation} often began with grandchildren proposing initial ideas, then turning to grandparents for narrative depth rooted in family memories. \textbf{Praise/Compliment} exchanges flowed both ways, with grandparents admiring grandchildren's technical skills, and grandchildren appreciating grandparents' creative suggestions. \textbf{Encouragement} typically flowed from grandparents to grandchildren, urging the younger participants to explore new interface functions or share story ideas. \textbf{Shared Laughter}, arising when discussing story ideas or playful story content, signaled emotional engagement. \textbf{Language Assistance} often saw grandchildren seeking help from grandparents for spelling or grammar checks. Conversely, \textbf{Clarification Requests} were common when grandchildren asked grandparents to interpret or explain the AI's responses, and vice versa, reversing the direction of help and balancing the relationship further. Lastly, \textbf{Technical Assistance} scenarios were not one-sided: while grandchildren usually led the digital interactions, they sometimes relied on grandparents to handle tasks like typing when they were unfamiliar using a keyboard. 
Through ongoing adjustments, participants constructed a shared creative space that respected generational differences as opportunities to learn from one another. This fluid interplay of roles helped transform technological and narrative gaps into bonding experiences.
}

\subsubsection{\revision{Encouraging Self-Reflection}}
\revision{
The AI-assisted co-creation activity fostered a reciprocal appreciation between generations. Grandchildren realized the value of their grandparents' storytelling abilities, discovering that narrative insight and life experience could meaningfully enrich the creative process. Grandparents realized that technology when navigated with the support of their grandchildren could feel less intimidating and more accessible.

One grandparent described how working alongside a grandchild reduced initial uncertainty, 
\textit{3GP0801: ``Oh it was great. Because of his experience, it would have taken me I think a lot more trying to get up to speed. The basic stuff. I think maybe it's partly age or life experience, but just the how do I do this? No. Intimidation, or computer programming. It helps that humans there.''}
This trust and guidance enabled grandparents to adopt new tools without feeling overwhelmed, reframing the learning process as collaborative rather than solitary. Also, grandchildren gained a heightened respect for their grandparents' adaptability and narrative contributions. Witnessing their grandparents engaging the AI-assisted co-creation activity challenged age-based assumptions about technology use.
} 
For example, one grandchild was inspired by their grandparent's adaptability (\textit{3GC0601, 3GP0601}), expressing warmth and admiration at the realization that older adults could find enjoyment in modern digital activities. Another grandchild reflected on how working together made the process more efficient and pleasant, stating, \textit{3GC0901: ``Yeah, we powered through that really quickly. We were coming up with ideas super quickly and working together very well.''} 
\revision{
These reflections underscore how AI-assisted co-creation activities bridged generational gaps, enabling each participant to appreciate the others' competencies, perspective and adaptability with new technologies. In doing so, the activity not only produced  creative outputs but also created an opportunity for self-reflection, further strengthening the intergenerational bonds.
}

\section{Discussion}
We explored how AI could facilitate co-creation and enhance communication between generations through a story creation activity during a community outreach event. In our study, 67 participants, grouped into 29 grandparent-grandchild groups, engaged in an AI-supported story creation task using \oursys. \revision{We aimed to answer the following research questions: (1) How do participants from different generations perceive AI's role within co-creation activities?; 
(2) In what ways do AI-supported co-creation activities facilitate intergenerational communication and connection?}

\revision{
Our findings, described in Section \ref{sec:findings}, reveal that AI plays a significant role in facilitating intergenerational co-creation and enhancing communication. Participants perceived AI as a structural facilitator, providing organization and focus during the story creation process. It was also perceived as a brainstorming partner, sparking discussions and idea-sharing between grandparents and grandchildren. Furthermore, AI was perceived as a assistant that enhanced narrative quality by refining inputs and bridged technological gaps, enabling meaningful contributions from both generations despite their varying levels of digital literacy. These roles highlight how AI was viewed as a mediator, amplifying the complementary strengths of grandparents and grandchildren and fostering intergenerational cohesion.

However, generational differences in perceptions of AI emerged during the activity. While grandchildren often embraced AI as a creative partner, some grandparents expressed skepticism or discomfort. This skepticism underscores the importance of recognizing how varying levels of familiarity and trust in technology influence the user experience. Despite these concerns, the activity itself facilitated shifts in perception for some grandparents. Grandparents, guided by their grandchildren, grew more comfortable with AI, appreciating its role in supporting the creative process. This hands-on engagement proved instrumental in bridging generational gaps in AI acceptance.

Additionally, the study illuminated the synergy between generations during the co-creation process. Grandchildren's digital fluency enabled smooth interaction with AI, while grandparents enriched the storytelling process with their narrative depth and personal insights. This reciprocal dynamic fostered mutual appreciation, with grandchildren valuing their grandparents' storytelling expertise and grandparents gaining confidence in technology through their grandchildren's guidance. Such interactions created a space for both generations to reflect on their unique contributions, fostering a more harmonious experience.
}

In this section, we further discuss the implications of our findings and present design implications of how can AI be integrated into co-creation activities in intergenerational groups.


\subsection{Design Implications}

\subsubsection{Considering Intergenerational Roles When Customizing AI}
\revision{
While our findings suggest that AI's adaptive capabilities, such as tailoring explanations based on the user's generation, could improve usability, this approach risks diminishing the unique contributions each generation brings. Grandparents often provided life experiences, storytelling depth, and linguistic support, whereas grandchildren offered technical fluency. For example, we observed an interaction scenario \textbf{Clarification Requests} (see Table ~\ref{tab:interactions_human}) where a grandparent from S3G10 mediated between the AI and a grandchild when the grandchild was confused by the AI's response. Similarly, a grandparent from S1G02 required a grandchild to explain the functionality of a loading popup UI element when confused by the technology. 

Customizing AI to provide adaptive learning systems has been shown to be beneficial in educational contexts by adjusting content in real-time based on feedback, tailoring teaching methods to suit individual learning styles and speeds \cite{kolluru2018adaptive, pesovski2024generative}. However, while customization can enhance usability, overly adaptive AI might inadvertently ``take over'' roles that are crucial for maintaining intergenerational interaction. Based on our findings, when integrating AI for activities involving multiple generations, we suggest designers to carefully consider the extent of customization. The goal should be to support and enhance intergenerational roles without removing the unique contributions that each generation brings, thereby providing opportunities for interaction and bonding rather than overshadowing human expertise.
}

\subsubsection{Expanding Users' Mental Models of AI Capabilities}
\revision{
Despite the AI's diverse capabilities, most participants tended to view it primarily as a ``story support'' tool. Even when opportunities arose to utilize AI for other tasks—such as clarifying difficult words or providing spelling assistance—participants asked each other rather than fully leveraging the AI's broader functionality (see Table ~\ref{tab:interactions_human}). This behavior suggests that participants had a limited understanding of AI's potential beyond narrative assistance. This narrow conceptualization can be attributed to participants' limited prior experience with AI like ChatGPT, as evidenced by Figure ~\ref{fig:pre-survey}. Without explicit prompts or instructions encouraging exploration of additional capabilities, participants limited their interactions to story creation tasks. As a result, AI's impact on their co-creation efforts remained restricted to its perceived primary function. 


To address the limited mental models \cite{johnson1983mental,norman2014some} observed in our study, we encourage designers to integrate the concept of play for intergenerational participants. Playfulness has the potential to transform AI-supported activities to into a source of fun and challenge for users to explore the boundaries of the AI’s limitations \cite{villareale2021understanding}. Also, leveraging positive AI-supported co-creation experiences can shift perspectives and demystify AI \cite{queiroz2020ai}. 

}

\subsubsection{Encouraging Self-Reflection and Mutual Appreciation}
\revision{
AI-assisted story co-creation provided participants with opportunities to reflect on their collaborative processes and personal interactions. Grandchildren recognized the value of their grandparents' storytelling abilities, while grandparents appreciated the supportive role of technology facilitated by their grandchildren. For instance, grandparents felt more competent and less intimidated by technology when co-creating with grandchildren, as noted in \textit{3GP0801}. This aligns with findings from \cite{freeman2020intergenerational} which highlights enhancing intergenerational relationships and technological adoption.

To further promote self-reflection and mutual appreciation, AI systems can incorporate features that prompt users to evaluate their contributions and the collaborative process. For example, after completing a storytelling session, the AI could ask, ``What aspect of working together during activity did you find most rewarding?'' Additionally, providing visual summaries of each participant's contributions can reinforce mutual appreciation and highlight the complementary roles each generation plays. 
}

\subsubsection{Structuring Interaction through Interactive Prompts}
\revision{
Prior work has shown that structured interaction procedures can enhance rapport and foster positive relationships in child-robot interactions \cite{coninx2016towards}, particularly during key moments such as initial encounters \cite{lee2022unboxing}.
In line with existing literature, our findings underscore the importance of providing structure and guidance within AI-assisted co-creation environments. Features like ``Help me'' buttons and predefined milestones supported both generations by offering clear, incremental goals and manageable steps in the storytelling process. By leveraging traditional UI elements—buttons, prompts, and progress indicators—the system helped participants stay focused, and prevent them from feeling lost. However, the system must also be flexible based on user experience. A scaffolding process is needed to provide experienced users with greater flexibility, as participant 2GP0701, an experienced writer, expressed frustration with the slow progress of the activity when constrained by the milestone-based structure. This adaptive approach ensures that the system meets the diverse needs of intergenerational users, fostering meaningful \camera{collaboration}{co-creation} across varying skill levels.

We suggest designers to incorporate intuitive and familiar UI components to frame AI interactions as guided, step-by-step processes. Using structured prompts can encourage sustained dialogue between generations, reducing the likelihood of either group feeling overwhelmed or uncertain about how to proceed. For example, implementing features such as a ``Help me'' button that provides documentation can assist users in completing their tasks while clarifying system capabilities. Additionally, visualizing milestones as a checklist aligns with the ``Visibility of system status'' principle, ensuring users remain informed of their progress through timely and appropriate feedback \cite{9800084}.
}


\subsubsection{Enhancing Credibility through Explainability and Transparency}
\revision{
Participants' concerns about the AI's credibility, particularly among grandparents, underscore the importance of explainability. Instances of AI hallucinations led to confusion, as also reported in prior work \cite{ji2023survey}, which can lead to distrust and disuse \cite{lee2004trust}. Additionally, participants often attributed AI errors to their own mistakes as people revise their mental models most when anomalies persist \cite{gero2020mental}. To mitigate these issues, AI systems should prioritize explainability and transparency.
}

\revision{
A pivotal approach to calibrating human trust is to convey AI's capability to humans \cite{bansal2021does, 10.1145/3411764.3445315, turner2022calibrating, wang2021explanations}. Designers should provide tutorials and explanations that reflect AI capability and behaviors \cite{lai2020chicago, lai2019human, 10.1145/3411764.3445315, lee2024rex} and consider users' specific needs and personal backgrounds to enhance understanding and engagement \cite{weitz2021demystifying}. Designers could also show confidence scores to help calibrate people's trust in an AI model \cite{10.1145/3351095.3372852}. 
Furthermore, increasing transparency by revealing basic information about how the AI generates suggestions or referencing known story elements can alleviate credibility concerns. Simple interface cues—such as indicating when AI suggestions stem from previously provided content—can be effective. By enhancing transparency and explainability, designers can help maintain engagement and encourage both generations to rely on the AI's input as a supportive, rather than suspect, resource.
}

\subsubsection{Supporting, Not Replacing Humans}
During our study, we observed that participants preferred AI to act as a supportive tool rather than a replacement for human action. While AI facilitated the creative process by offering suggestions and guiding participants through the story creation, participants remained the primary decision-makers. Both grandparents and grandchildren appreciated this autonomy, using AI for structured support and inspiration when needed, but maintaining ownership of their choices during \camera{collaborative}{co-creation} activities.

\revision{
To ensure that AI supports rather than replaces human expertise, designers should implement soft intervention approaches where AI offers ideas without dictating the process. For instance, the AI could suggest, ``How about introducing a conflict between Tom and Steve?'' while allowing users to accept or modify the suggestion. Additionally, providing options for users to control the level of AI assistance ensures that human expertise remains central to the \camera{collaboration}{activities} \cite{norman2009design, stegner2024understanding, westphal2023decision, 8160794}. 

}

\subsection{Limitations and Future Work}
Our work has a number of limitations. First, the occurrence of LLM hallucinations led to AI generating incorrect or confusing story elements, which disrupted the narrative of the story. Future work should focus on refining error-handling mechanisms to prevent such disruptions and ensure a smoother co-creation experience.
Second, our study was conducted as a one-time workshop experience during a community outreach event, which may limit the depth of insight into the long-term impact of AI-assisted co-creation. Future work should explore longer-term deployments to assess how sustained use of AI in intergenerational activities might evolve over time and further shape participants’ experiences and perceptions.
Third, our study primarily involved participants from an outreach community event, potentially limiting the generalizability of the findings to populations, where digital literacy and access to technology may differ. Future work should include participants from a wider range of geographical and socioeconomic backgrounds to better understand how AI can be tailored to diverse intergenerational contexts.

\section{Conclusion}
In this paper, we conducted a study with 29 grandparent-grandchild groups, exploring how AI-assisted co-creation could facilitate intergenerational \camera{collaboration}{co-creation} in a workshop setting. Participants engaged in a story creation activity supported by an AI system, where we observed that grandchildren typically managed the technical aspects while grandparents contributed creatively to guide the co-creation process. 
\revision{
This complementary distribution of tasks, enabled by the AI’s structured prompts and refinements, facilitated a smoother, more cohesive co-creation experience.
Our findings demonstrate that AI can serve as a catalyst for bridging generational differences in technological familiarity, fostering richer dialogue and mutual appreciation. While grandchildren embraced AI as a resourceful collaborator, grandparents approached it with more caution, highlighting the importance of transparency, explainability, and balanced support. These insights underscore the potential for AI-assisted co-creation to enhance intergenerational \camera{collaboration}{bonds}, provided that systems are designed to acknowledge both generations' strengths and needs.
}
Based on these insights, we present design implications for AI systems that support intergenerational co-creation, emphasizing the need for an inclusive and adaptable system that caters to varying levels of technological familiarity and skill.

\begin{acks}
This work was in part supported by the National Science Foundation awards 1925043, 2312354, and 2152163 as well as the Sheldon B. and Marianne S. Lubar Professorship. Figures \ref{fig:teaser} and Figure \ref{fig:findings} include images by \texttt{upklyak}, \texttt{redgreystock}, \texttt{Talha Dogar}, \texttt{iconixar} and \texttt{storyset} on Freepik. We thank Debrah Fosu and Taenam Kim for their assistance with data analysis and Yaxin Hu for her support in data collection.
\end{acks}

\bibliographystyle{ACM-Reference-Format}
\bibliography{reference}

\appendix
\section{Technology Familiarity Questionnaire}
\label{sec:pre-survey}

To measure participants familiarity with technology, we used the following questions:

\begin{enumerate}
    \item Have you ever used ChatGPT before? (Yes/No)
    \item Do you have any programming experience? (Yes/No)
    \item[ ] If ``Yes:''
    \begin{enumerate}
        \item How many years of programming experience do you have?
        \begin{itemize}
            \item Less than 6 months
            \item 6 months to 1 year
            \item 1-2 years
            \item 3-5 years
            \item More than 5 years
        \end{itemize}
        \item What kind of programming languages have you used?
        \begin{itemize}
            \item Python
            \item Java
            \item JavaScript
            \item C++
            \item Scratch
            \item Other (please specify)
        \end{itemize}
        \item Have you ever programmed a robot? (Yes/No; if ``Yes,'' please describe your experience.)
    \end{enumerate}
\end{enumerate}

\section{Interview}
\label{sec:interview}
The interview questions included:

\begin{enumerate}
    \item Tell us about your experience with Roboblock (Follow-up questions: why, what else)
    \begin{enumerate}
        \item Did grandparents help grandchildren using the system?
    \end{enumerate}
    \item Did grandparents help grandchildren using the system
    \item What are your thoughts on using AI for story generation and programming?
    \item Were there any things you wanted to do with Roboblock but could not?
    \item What parts of Roboblock appealed to you the most? (must be asked to at least one grandparent and one grandchild to explore any differences in appeal between generations) 
    \item Can you think of ways in which Roboblock can be improved? 
\end{enumerate}

\end{document}